\documentclass[iop]{emulateapj}
\usepackage{natbib}
\usepackage{epsfig}
\usepackage{graphicx}
\usepackage{epstopdf}

\newcommand\icarus{{Icarus}}
\newcommand{\Rpl}{R_{\rm{pl}}}
\newcommand{\Mpl}{M_{\rm{pl}}}
\newcommand{\tdrift}{t_{\rm{drift}}}
\newcommand{\rau}{r_{\rm{AU}}}
\newcommand{\tfric}{t_{\rm{fric}}}
\newcommand{\taufric}{\tau_{\rm fric}}
\newcommand{\unitkappa}{{\rm cm^{2}\:g^{-1}}}
\newcommand{\sigmasb}{\sigma_{\rm{SB}}}
\newcommand{\phideltar}{\phi_{\Delta r}}
\newcommand{\phideltari}{\phi_{\Delta r,i}}
\newcommand{\deltar}{\Delta r}
\newcommand{\deltari}{\Delta r_i}
\newcommand{\phideltarone}{\phi_{\Delta r,1}}
\newcommand{\phideltartwo}{\phi_{\Delta r,2}}
\newcommand{\phideltarthree}{\phi_{\Delta r,3}}
\newcommand{\phideltarfour}{\phi_{\Delta r,4}}
\newcommand{\Mgap}{M_G}
\newcommand{\gccc}{\rm{gcm^{-3}}}
\newcommand{\kepler}{{\it Kepler}}

\newcommand     \He     {{\rm He}}
\newcommand     \HH     {{{\rm H}_2}}

\newcommand     \kms    {\,{\rm km~s}^{-1}}

\newcommand     \s      {\,{\rm s}}

\newcommand     \yr     {\,{\rm yr}}

\newcommand{\smyr}{{ M_\odot\ \rm yr^{-1}}}

\newcommand{\beq}{\begin{equation}}
\newcommand{\eeq}{\end{equation}}
\newcommand{\beqa}{\begin{eqnarray}}
\newcommand{\eeqa}{\end{eqnarray}}

\newcommand{\gcc}         {\rm g\:cm^{-2}}

\newlength{\figwidth}
\shorttitle{Inside-Out Planet Formation}
\shortauthors{Chatterjee \& Tan}

\begin{document}

\title{Inside-Out Planet Formation}

\author{Sourav Chatterjee}
\affil{Department of Astronomy, University of Florida, Gainesville, FL 32611, USA\\s.chatterjee@astro.ufl.edu}
\author{Jonathan C. Tan}
\affil{Departments of Astronomy \& Physics, University of Florida, Gainesville, FL 32611, USA\\jt@astro.ufl.edu}

\begin{abstract}
The compact multi-transiting planet systems discovered by {\it Kepler}
challenge planet formation theories. Formation {\it in situ} from disks
with radial mass surface density, $\Sigma$, profiles similar to the
minimum mass solar nebula (MMSN) but boosted in normalization by
factors $\gtrsim10$ has been suggested.  We propose that a more
natural way to create these planets in the inner disk is formation
sequentially from the inside-out via creation of successive
gravitationally unstable rings fed from a continuous stream of small
($\sim$cm--m size) ``pebbles'', drifting inwards via gas drag. Pebbles
collect at the pressure maximum associated with the transition from a
magneto-rotational instability (MRI)-inactive (``dead zone'') region
to an inner MRI-active zone. A pebble ring builds up until it either
becomes gravitationally unstable to form an $\sim$1~$M_\Earth$ planet
directly or induces gradual planet formation via core accretion. The
planet may undergo Type I migration into the active region, allowing a
new pebble ring and planet to form behind it. Alternatively if
migration is inefficient, the planet may continue to accrete from the
disk until it becomes massive enough to isolate itself from the
accretion flow. A variety of densities may result depending on the
relative importance of residual gas accretion as the planet approaches
its isolation mass. The process can repeat with a new pebble ring
gathering at the new pressure maximum associated with the retreating
dead zone boundary. Our simple analytical model for this scenario of
inside-out planet formation yields planetary masses, relative mass
scalings with orbital radius, and minimum orbital separations
consistent with those seen by {\it Kepler}. It provides an explanation
of how massive planets can form with tightly-packed and well-aligned
system architectures, starting from typical protoplanetary disk
properties. 
\end{abstract}

\section{Introduction}\label{S:intro}

A striking property of the {\it Kepler}-detected planet candidates
(KPC) is the existence of multi-transiting systems with tightly-packed
inner planets (STIPs): typically 3--5 planets of radii
$\sim1-10\:R_\oplus$ in short-period (1--100d) orbits
\citep{2012ApJ...761...92F}.  While short-period giant planets can be
explained via planet-planet scattering followed by tidal
circularization
\citep{1996Sci...274..954R,2008ApJ...686..580C,2011ApJ...742...72N},
this mechanism cannot produce the low dispersion ($\lesssim3^\circ$)
in orbital inclinations of STIPs. 
Their well-aligned orbits imply either formation {\it in situ}
within a disk or formation at larger distances followed by inward
migration within a gas disk. The migration scenario has been discussed
by, for example, 
\citet{2012ARA&A..50..211K}: it tends to produce planetary orbits that are
trapped near low-order mean motion resonances. However, such pile-ups of orbits
near resonances do not appear to be a particular feature of the KPCs,
so other mechanisms would then be needed to move planets away from
resonance \citep[e.g.,][]{2012ApJ...756L..11L,2012MNRAS.427L..21R,2013AJ....145....1B}. 

Formation {\it in situ} faces the problem of concentrating
a large mass of solids in the inner disk.  \citet{2013MNRAS.431.3444C}
used the observed distribution of KPCs to construct a $\Sigma$ profile
of a typical disk that would form such planets, finding it has
significantly more solids within $\sim1\:\rm{AU}$ than the MMSN. They
then discussed several implications of forming planets from such a
disk.  \citet{2012ApJ...751..158H,2013arXiv1301.7431H} proposed this
concentration ($\sim 20\:M_\oplus$ inside 1~AU) is achieved via
migration of small bodies to form an inner enriched disk. They then
considered a standard model for planet formation via oligarchic growth
from such a disk.

Here we present an alternative model involving {\it simultaneous}
migration of small ($\sim$cm--m) solids (hereafter ``pebbles"), and
planet formation at the location where these pebbles are
deposited. Inward migration of pebbles occurs via gas drag due to the
disk's radial pressure gradient --- long recognized as part of the
so called ``meter-size barrier'' for planetesimal formation
\citep{1977MNRAS.180...57W,2013pss3.book....1Y}.
However, although this inhibits planet formation in most of the disk,
we argue it is key for enabling close-in massive planet formation. 

In Section\ 2 we describe our proposed scenario and present a
  simple analytical model to calculate the predicted planetary masses,
  mass--orbital distance relation, and minimum planet-planet
  separations. In Section\ 3 we compare the predicted planetary
  properties with the observed \kepler\ systems. Finally, in
  Section\ 4 we summarize and discuss implications of our model, as
  well as identifing caveats that can be tested via future
  numerical simulations.

\section{Overview of Theoretical Model}
\label{S:theory}

\begin{figure*}
\begin{center}
\plotone{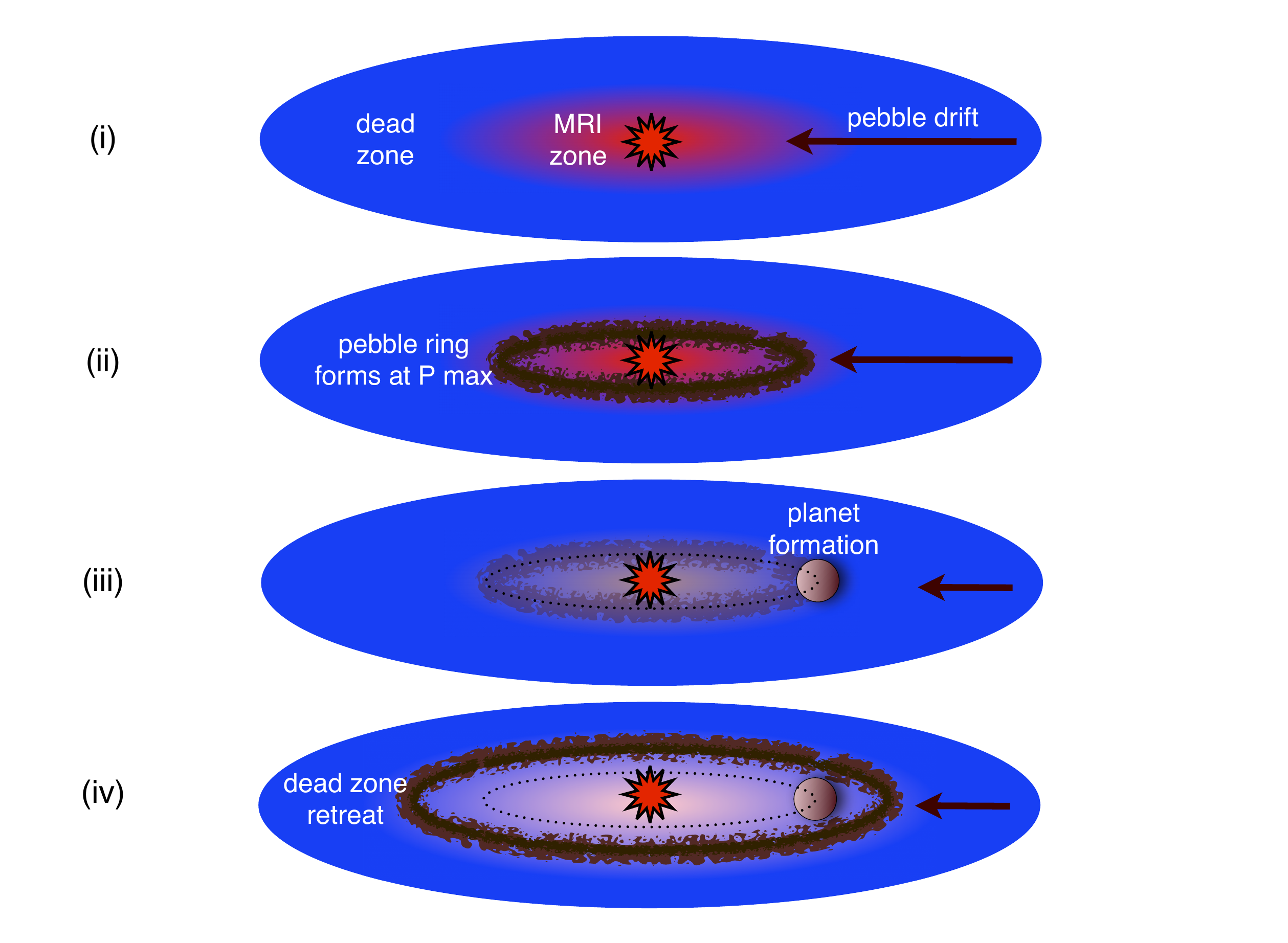}
\caption[$M$ vs $r$]{
Schematic overview of the stages of the inside-out planet formation
scenario. (i) Pebble formation and drift to the inner disk. Pebbles
form via dust coagulation in the protoplanetary disk. Those with
$\sim$cm--m sizes attain high radial drift velocities and quickly
reach the dead zone inner boundary, where they become trapped at the
pressure maximum. (ii) Pebble ring formation. A ring of pebbles
gradually builds up over a timescale set by the pebble formation and
supply rate from the outer disk. (iii) Planet formation and gap
opening. A planet forms either via gravitational (Toomre) instability
of the ring or via core accretion. In both cases, a gradual
accumulation of the bulk of the ring mass into a single planet is
anticipated. In the case of gravitational instability, this ring mass,
once organized into a single planet, may be larger than the mass
needed to open a gap in the gas disk. For core accretion, the final
planet mass may be limited by such gap opening. In both cases, gap
opening is soon followed by viscous clearing of the gas disk interior
to the planet's orbit. (iv) Dead zone retreat and subsequent pebble
ring and planet formation.  Gap opening and associated viscous
clearing of the inner disk allow greater penetration of X-ray photons
from the protostar to the disk mid-plane, increasing its ionization
fraction and thus activating the magneto-rotational instability
(MRI). The inactive dead zone retreats, along with the pressure
maximum associated with its inner boundary. A new pebble ring starts
to form at this location that can form a new planet. This cycle
repeats leading to sequential formation of a planetary system from the
inside-out.
}
\label{fig:cartoon}
\end{center}
\end{figure*}

A schematic overview of the model is presented in
Figure~\ref{fig:cartoon}, involving four basic stages: (i) Pebble
formation and drift to the inner disk; (ii) Pebble ring formation at
the pressure maximum associated with dead zone inner boundary; (iii)
Planet formation from the pebble ring leading to gap opening and
viscous clearing of the inner disk; (iv) Dead zone retreat and
formation of a new pebble ring that can lead to subsequent planet
formation. These stages are described in more detail below.

\subsection{Pebble Formation and Drift to the Inner Disk}\label{S:pebble_form}

Consider an accretion disk of total mass $M$, composed of gas
($M_{g}$) and solids ($M_s$). We class solids in two types: (1)
$\lesssim$sub-mm dust grains ($M_{d}$), perfectly coupled to gas; (2)
$\gtrsim 1\:\rm{cm}$ ``pebbles'' ($M_{p}$) that feel significant gas
drag.  Thus $M_s=M_d+M_p$. The disk is remnant material from star
formation with interstellar composition, i.e., $M_s=f_sM_g$ with
$f_s\simeq0.01$.

We will consider a Shakura-Sunyaev alpha-disk model for the
protoplanetary disk \citep{1973A&A....24..337S}. The viscosity parameter
$\alpha\equiv0.001\alpha_{-3}=\nu/(c_sH)$, is expected to be
$\sim10^{-2}$ in MRI-active regions, but at least an order of
magnitude smaller in the dead zone, where the ionization fraction is
too small for the magnetic field to couple to the gas. For example,
\citet[][]{2010A&A...515A..70D} find $\alpha \sim10^{-4}-10^{-3}$ set
by diffusion of the mean magnetic field from the active region into
the dead zone. Values of $\alpha$ of this order may also result from
damping and shocking of spiral density waves that are excited by a
planet that is already present in the inner region of the disk
\citep{2001ApJ...552..793G}. While MRI-active disk surface layers may be
expected above and below the dead zone (although even here,
suppression of the MRI by the global magnetic fields associated with a
disk wind remains a possibility; G. Lesur, private communication), here we are
concerned with mid-plane properties, where pebbles will have settled.

For a steady, thin, active accretion disk, the mid-plane properties such 
as pressure ($P$), temperature ($T$), sound speed ($c_s$) can be derived 
as a function of the basic disk properties including the accretion rate ($\dot{m}$), 
disk composition, opacity, and assumed $\alpha$ viscosity \citep[for a detailed discussion and 
derivation of the structure of alpha-disks see, e.g.,][]{2002apa..book.....F}. Following 
closely the treatment in \citet{2002apa..book.....F}, we can write the mid-plane pressure 
as  
\begin{eqnarray}
\label{eq:pressure}
P&=&\frac{2^{1/2}}{3^{11/10}\pi^{4/5}}\left(\frac{\mu}{k_B}\right)^{2/5} \gamma^{-7/5} \left(\frac{\kappa}{\sigmasb}\right)^{-1/10}\alpha^{-9/10}\nonumber\\
&\times&(G m_*)^{17/20}(f_r\dot{m})^{4/5}r^{-51/20},\\
P/k_B&\rightarrow&1.22\times10^{16}\gamma_{1.4}^{-7/5} \kappa_{10}^{-1/10}\alpha_{-3}^{-9/10}\nonumber\\
&\times&m_{*,1}^{17/20}(f_r\dot{m}_{-9})^{4/5}\rau^{-51/20}\:{\rm{K\:cm^{-3}}}\nonumber
\end{eqnarray}
where $\mu=2.33m_{\rm{H}}=3.90\times10^{-24}\:\rm{g}$ is the mean
particle mass (assuming $n_\He=0.2n_\HH$), $k_B$ is Boltzmann's
constant, $\gamma \equiv 1.4\gamma_{1.4}$ is the power law exponent of
the barotropic equation of state $P=K\rho^\gamma$ where we have
normalized for $\rm H_2$ with rotational modes excited,
$\sigmasb$ is Stefan-Boltzmann's constant, $m_* \equiv m_{*,1}
M_\odot$ is the stellar mass,
$\kappa \equiv  \kappa_{10} 10\:\unitkappa$ is disk opacity
\citep[normalized to expected protoplanetary disk values,
  e.g.,][]{2002ApJ...564..887W}, $f_r\equiv1-\sqrt{r_*/r}$, (where
$r_*$ is stellar radius), and
$\dot{m}\equiv \dot{m}_{-9} 10^{-9}\:\smyr$ is the accretion rate.
We have normalized $\dot{m}$ to expected protoplanetary disk values,
although these show wide dispersion and may also individually vary
over time, i.e., possible accretion bursts superposed on longer term
decline \citep[e.g.,][]{2011ARA&A..49...67W}.

Similarly, the density, $\rho = \gamma P c_s^{-2}$, in the disk
  mid-plane is given by
\begin{eqnarray}
\label{eq:density}
\rho&=&\frac{2^{3/2}}{3^{13/10}\pi^{2/5}}\left(\frac{\mu}{\gamma k_B}\right)^{6/5} \left(\frac{\kappa}{\sigmasb}\right)^{-3/10}\alpha^{-7/10}\nonumber\\
&\times&(G m_*)^{11/20}(f_r\dot{m})^{2/5}r^{-33/20},\\
 &\rightarrow&1.87\times10^{-10}\gamma_{1.4}^{-6/5}\kappa_{10}^{-3/10}\alpha_{-3}^{-7/10}\nonumber\\
&\times&m_{*,1}^{11/20}(f_r\dot{m}_{-9})^{2/5}\rau^{-33/20}\:{\rm{g\:cm^{-3}}}\nonumber
\end{eqnarray}
(where this fiducial density corresponds to a number density of $\rm
H_2$ molecules of $n_{\rm H_2} = 4.00\times 10^{13}\:{\rm cm^{-3}}$). 
It then follows that the disk mid-plane sound speed is
\begin{eqnarray}
\label{eq:cs}
c_s&=&\frac{3^{1/10}}{2^{1/2}\pi^{1/5}}
\left(\frac{\mu}{\gamma k_B}\right)^{-2/5} 
\left(\frac{\kappa}{\sigmasb}\right)^{1/10}
\alpha^{-1/10}\nonumber\\
&\times& \left(f_r\dot{m}\right)^{1/5} (Gm_*)^{3/20} r^{-9/20}\\
&\rightarrow&1.12\kappa_{10}^{1/10}\gamma_{1.4}^{2/5}\alpha_{-3}^{-1/10}m_{*,1}^{3/20}(f_r\dot{m}_{-9})^{1/5}\rau^{-9/20}\:\kms,\nonumber
\end{eqnarray}
and the disk mid-plane temperature is
\begin{eqnarray}
\label{eq:temp}
T&=&\frac{3^{1/5}}{2\pi^{2/5}}\left(\frac{\mu}{\gamma k_B}\right)^{1/5} \left(\frac{\kappa}{\sigmasb}\right)^{1/5}\alpha^{-1/5}\nonumber\\
&\times&(G m_*)^{3/10}(f_r\dot{m})^{2/5}r^{-9/10},\\
 &\rightarrow&254\gamma_{1.4}^{-1/5}\kappa_{10}^{1/5}\alpha_{-3}^{-1/5}m_{*,1}^{3/10}(f_r\dot{m}_{-9})^{2/5}\rau^{-9/10}\:{\rm K}.\nonumber
\end{eqnarray}

Disk solids grow from dust grains to pebbles at rate
$\dot{M}_p=-\dot{M}_d$ set by coagulation of small grains into larger
ones---a complicated process expected to depend on grain structure and
composition \citep[e.g.,][]{2010RAA....10.1199B}.  There is thus a
radially-varying source term for pebbles, dependent on dust grain
number density. We will see later that we will mostly be concerned
with the supply of pebbles forming at locations $r\sim 10$~AU.

A decreasing pressure gradient in the disk (Equation~\ref{eq:pressure}) causes gas to orbit at
slightly sub-Keplerian speeds. Pebbles of mass $m_p$ and radius $a_p\equiv a_{p,1}$~cm,
whose orbits are not affected by this pressure gradient due to low coupling with gas, have
relative velocities with respect to the gas of magnitude 
$v_\Delta \simeq t_{\rm fric} v_{r,p} v_K / (2 r)$ 
\citep{1977MNRAS.180...57W,2002ApJ...581.1344T,2007astro.ph..1485A}. 
Here $t_{\rm fric}$ is the frictional timescale, $v_{r,p}$ is the radial
drift speed of the pebble and $v_K$ is the Keplerian speed. The
frictional time scale is defined as $t_{\rm fric} \equiv m_p v_\Delta
/ |F_D|$, where $F_D$ is the drag force. The drag force is $|F_D| =
(1/2) C_D \pi a_p^2 \rho v_\Delta^2$, where $C_D$ is the drag
coefficient. The Epstein regime of drag applies when
$a_p<(9/4)\lambda$, where $\lambda$ is the mean free path of
molecules, given by
\begin{eqnarray}\label{lambda}
\lambda & = & \frac{1}{n_{\rm H2} \sigma_{\rm H2}}\\
 & = & 12.5 \gamma_{1.4}^{-6/5} \kappa_{10}^{3/10} \alpha_{-3}^{7/10} m_{*1}^{-11/20}(f_r\dot{m}_{-9})^{-2/5} r_{\rm AU}^{33/20}\:{\rm cm},\nonumber
\end{eqnarray}
where we have adopted $\sigma_{\rm H2}=2\times 10^{-15}\:{\rm
  cm^2}$. In this regime $C_D = 2^{9/2} c_s/ (3\sqrt{\pi}v_\Delta)$.
As discussed by \citet{2007astro.ph..1485A}, the pebble inward radial
drift velocity $v_{r,p}$, depends on pebble size and the
disk's pressure profile $P=P_0(r/r_0)^{-k_P}$, and is given by
\begin{eqnarray}
\label{eq:vr}
v_{r,p} & \simeq & \frac{-k_P(c_s/v_K)^2}{\taufric+\taufric^{-1}} v_K \nonumber\\
|v_{r,p}| & \simeq & \frac{3^{1/5}}{2\pi^{2/5}} f_\tau k_P \left(\frac{\mu}{\gamma k_B}\right)^{-4/5}
\left(\frac{\kappa}{\sigmasb}\right)^{1/5}
\alpha^{-1/5}\nonumber\\
&\times&(G m_*)^{-1/5}
(f_r\dot{m})^{2/5}
r^{-2/5}\\
&\rightarrow & 0.108 f_\tau \gamma_{1.4}^{4/5}\kappa_{10}^{1/5}\alpha_{-3}^{-1/5}m_{*,1}^{-1/5}\nonumber\\
 & \times & (f_r\dot{m}_{-9})^{2/5}\rau^{-2/5}\:\kms.\nonumber
\end{eqnarray}
where 
$\taufric\equiv\Omega_K\tfric$ is the normalized
pebble frictional time, where $\Omega_K=(GM/r^3)^{1/2}$, and $f_\tau \equiv (\taufric+\taufric^{-1})^{-1}$.
For the alpha disk given by Equation~\ref{eq:pressure}, $k_P=51/20=2.55$. 

The radial drift timescale is then 
\begin{eqnarray}
\label{eq:tdrift}
\tdrift& \equiv & \frac{r}{|v_{r, p}|}\nonumber\\
 &=&\frac{2\pi^{2/5}}{3^{1/5}f_\tau k_P}
\left(\frac{\mu}{\gamma k_B}\right)^{4/5}
\left(\frac{\kappa}{\sigmasb}\right)^{-1/5} \alpha^{1/5}
\nonumber\\
&\times&  (Gm_*)^{1/5}
(f_r\dot{m})^{-2/5} r^{7/5}\\
&\rightarrow&43.9f_\tau^{-1} \gamma_{1.4}^{-4/5}\kappa_{10}^{-1/5}\alpha_{-3}^{1/5}m_{*,1}^{1/5}(f_r\dot{m}_{-9})^{-2/5} r_{\rm{AU}}^{7/5}\:\yr.\nonumber
\end{eqnarray}
If $\taufric \sim {\cal O}(1)$, then $\tdrift$ for pebbles is much shorter than the disk lifetime, expected to be $\gtrsim1$~Myr \citep{2011ARA&A..49...67W}.

Growing from small dust grains, pebbles will first be in the Epstein
drag regime. In this case
\begin{eqnarray}
\label{eq:taufric}
\taufric &=& \frac{3^{6/5}\pi^{11/10}}{2^{5/2}} \rho_{p} a_p
\left(\frac{\mu}{\gamma k_B}\right)^{-4/5}
\left(\frac{\kappa}{\sigmasb}\right)^{1/5}
\alpha^{4/5}\nonumber\\
&\times&(G m_*)^{-1/5}
(f_r\dot{m})^{-3/5}
r^{3/5}\\
&\rightarrow&0.0178 a_{p,1} \rho_{p,3} \gamma_{1.4}^{4/5}\kappa_{10}^{1/5}\alpha_{-3}^{4/5}m_{*,1}^{-1/5}(f_r\dot{m}_{-9})^{-3/5}\rau^{3/5},\nonumber
\end{eqnarray}
where $\rho_p\equiv \rho_{p,3}3\:{\rm g\:cm^{-3}}$ is
the pebble density. In this limit where $\taufric \ll 1$, $f_\tau
\rightarrow \taufric$. However, the process of radial drift of pebbles
should lead to their rapid growth as they sweep up smaller dust
grains, especially in colder outer regions of the disk where grains
still retain their ice mantles. In the CQ~Tau protoplanetary disk,
Trotta et al. (2013) have derived maximum ``grain'' sizes of a few mm
at 80~AU, increasing to a few cm in the inner 40~AU. If growth is
efficient, then pebbles may reach sizes associated with the maximum
radial drift speed, i.e. for $\taufric =
1$. Equation~\ref{eq:taufric} can be rearranged to yield an
expression for pebble size:
\begin{eqnarray}
\label{eq:pebble_size}
a_{\rm p} &=& \frac{2^{5/2}}{3^{6/5}\pi^{11/10}}\frac{\tau_{\rm fric}}{\rho_{p}}
\left(\frac{\mu}{\gamma k_B}\right)^{4/5}
\left(\frac{\kappa}{\sigmasb}\right)^{-1/5}
\alpha^{-4/5}\nonumber\\
&\times&(G m_*)^{1/5}
(f_r\dot{m})^{3/5}
r^{-3/5}\\
&\rightarrow&56.2 \rho_{p,3}^{-1}\tau_{\rm fric}\gamma_{1.4}^{-4/5}\kappa_{10}^{-1/5}\alpha_{-3}^{-4/5}\nonumber\\
& \times & m_{*,1}^{1/5}(f_r\dot{m}_{-9})^{3/5}\rau^{-3/5}\:{\rm cm},\nonumber
\end{eqnarray}
The condition to be in the Epstein drag regime, $4a_p/(9\lambda)<1$,
can be evaluated via
\begin{equation}
\frac{4a_p}{9\lambda} = 1.99 \rho_{p,3}^{-1}\tau_{\rm fric}\gamma_{1.4}^{2/5}\kappa_{10}^{-1/2}\alpha_{-3}^{-3/2}m_{*,1}^{3/4}(f_r\dot{m}_{-9})\rau^{-9/4}.
\end{equation}
We see that for the fiducial disk parameters it is satisfied for
$\rau>1.36 \taufric^{4/9}$, i.e. for the bulk of the feeding zone that we expect to
be relevant for planet formation (\S\ref{S:subsequent}).  

Thus pebbles, growing from small sizes, will first be in the Epstein
drag regime and will drift inwards on relatively short timescales. As
they enter denser regions of the disk, they will enter the Stokes drag
regime, but continue drifting inwards.  While pebbles that have
$\taufric=1$ reach the inner disk the fastest, given the generally
short values of $t_{\rm drift}$, we expect the delivered material to
actually have a relatively broad distribution of sizes and thus also
radial drift speeds.
We conclude, like \citet{2012ApJ...751..158H}, that radial pebble
migration can provide a large reservoir of solids to build inner short
period planets. 
Given the short drift timescales, the rate limiting step for
the supply of pebbles to the inner dead zone boundary is likely to be
their formation rate via dust coagulation in the outer disk.

\subsection{Pebble Ring Formation at Inner Disk Pressure Maximum}
\label{S:pmax}

We assume there is an inner-disk location, $r_0$, where, moving
inwards, gas pressure declines rapidly from a local maximum, leading
to accumulation of pebbles.
We expect the mechanism responsible for initially creating this
central ``pressure hole'' is the transition from an outer
MRI-inactive, dead zone, region to an inner active region. We see from
Equation~\ref{eq:pressure}, that the pressure scales as
$P\propto\alpha^{-9/10}$, so as $\alpha$ rapidly increases on leaving
the inner dead zone boundary, mid-plane pressure decreases almost as
rapidly.
Note that although $\kappa$, set by dust opacity, is a function of
local disk properties (i.e. density and temperature), it is not
expected to vary strongly in this transition region. Moreover, $P$ has
a very weak dependence on $\kappa$. 
This analytical expectation for the existence of a pressure maximum
associated with the inner dead zone boundary is also seen in the results 
from numerical simulation by \citet{2010A&A...515A..70D}.

Depending on disk properties such as $\dot{m}$ and $\alpha$ and
stellar X-ray luminosity, $L_X$, the location of the inner dead zone
boundary in the disk mid-plane, i.e. the location where the ionization
fraction reaches a certain critical value, is likely to be set either
by thermal ionization of alkali metals at $T\sim 1200$~K
\citep{1988PThPS..96..151U} or by penetration of protostellar X-rays
that are produced from flares associated with magnetic activity both
near the stellar surface and possibly also from a disk corona. 
The X-ray luminosity of young stars is thus,
by its nature, highly variable and so the radial location of
the dead zone inner boundary could also fluctuate, depending on the
timescale for MRI turbulence to develop in response to a change in
ionization.

From Equation~\ref{eq:temp}, we see that the temperature of
$1200$~K for thermal ionization of alkali metals is achieved at
\begin{equation}
\label{eq:rinner}
r_{1200{\rm K}}=0.178\gamma_{1.4}^{-2/9}\kappa_{10}^{2/9}\alpha_{-3}^{-2/9}m_{*,1}^{1/3}(f_r\dot{m}_{-9})^{4/9}\:{\rm AU}.
\end{equation}
If there is efficient extraction of accretion power as mechanical
luminosity of a disk wind, then the disk at a given radius will be cooler than predicted
by Equation~\ref{eq:temp}. For example, in the disk models of 
\citet{2013ApJ...766...86Z} including this effect of disk winds causes a given temperature zone to
be about 20\% closer to the star compared to models without
winds. 
These models also show that as $\kappa$ begins to decrease at
$T\gtrsim 1400$~K, the disk temperature is kept relatively constant
over a factor of several in radius. Given the uncertainties in dust
composition and the temperature of dust destruction, it is possible
that opacity reduction may begin at $T\simeq 1200$~K, thus keeping the
disk at this temperature to radii that are factors of several smaller
than predicted by Equation~\ref{eq:rinner}.

More detailed calculations of the thermal and ionization structure of
protostellar accretion disks, together with input models for the
global magnetic field structure, are needed for accurate prediction of
the location of the dead zone inner boundary due to thermal and X-ray
ionization.  Example calculations of dead zone boundaries have been
carried out by
\citet{2005ApJ...618L.137M,2010A&A...515A..70D,2013ApJ...764...65M,2013arXiv1305.1890O}.
For example, in the fiducial model of \citet{2013ApJ...764...65M}, the
dead zone extends inside 0.1~AU, but this is sensitive to model input
parameters. 

In summary, the dead zone inner boundary in an active accretion disk
is likely to be set by thermal ionization of alkali metals, and can be
at a small fraction of an AU, depending on the accretion rate. The
estimate given by Equation~\ref{eq:rinner} for $r_{\rm{1200K}}$, should
be regarded as an upper limit, since, both disk
wind energy extraction and opacity reduction due to dust destruction, act to reduce 
the disk mid-plane temperature compared to the value given in Equation\ \ref{eq:temp} 
at a given location.

Efficient pebble drift from the outer disk together with the strong
theoretical expectation of an inner local pressure maximum at the dead
zone inner boundary make it likely that a pebble ring will form at
this location.  We expect and will assume that the global
radial drift of pebbles, through the dead zone, to $r_0$ will
overwhelm any mechanism that may be acting to limit the concentration
of solids, such as vertical shear instabilities 
\citep[e.g.,][]{1980Icar...44..172W,2002ApJ...580..494Y}, turbulence induced 
by streaming instabilities \citep{2010ApJ...722L.220B,2010ApJ...722.1437B}, 
or Rossby wave instabilities \citep[e.g.,][]{2012A&A...545A.134M,2012ApJ...756...62L}.

Now we examine the condition that must be satisfied to ensure that after pebbles 
are delivered to $r_0$, they are trapped at the pressure maxima, instead of being carried 
further inwards by the radial inward flow of gas.  
The positive pressure gradient associated with the dead zone inner edge will induce a
net outward radial drift velocity of pebbles with respect to gas, and this velocity needs 
to be larger than the inward radial
flow of gas due to viscous accretion, $v_{r,g}$, given by
\begin{eqnarray}
\label{eq:vg}
v_{r,g} & = & - 3 \nu / (2 f_r r)\nonumber\\
|v_{r,g}| & = & \frac{3^{6/5}}{4\pi^{2/5}} \left(\frac{\mu}{\gamma k_B}\right)^{-4/5}
\left(\frac{\kappa}{\sigmasb}\right)^{1/5}
\alpha^{4/5}\nonumber\\
&\times&(G m_*)^{-1/5}
f_r^{-3/5}\dot{m}^{2/5}
r^{-2/5}\\
&\rightarrow & 6.34 \gamma_{1.4}^{4/5}\kappa_{10}^{1/5}\alpha_{-3}^{4/5}m_{*,1}^{-1/5} f_r^{-3/5} \dot{m}_{-9}^{2/5}\rau^{-2/5}\:{\rm cm\: s^{-1}}.\nonumber
\end{eqnarray}
Describing the edge pressure gradient by a power law $\propto r^{-k_{P,\rm{edge}}}$ 
, we expect $k_{P,{\rm edge}}<-k_P \rightarrow
-2.55$ and will normalize to a fiducial value of -10. Then the
condition $v_{r,p}>|v_{r,g}|$ implies $\taufric > 3\alpha/(2 k_{P,{\rm
    edge}} f_r)$. In the Epstein drag regime with $\taufric \ll 1$,
then Equations~\ref{eq:taufric} and \ref{eq:rinner} imply
\begin{equation}
\label{eq:apmin}
a_p > 0.0237 \rho_{p,3}^{-1}k_{P,{\rm edge},-10}^{-1}\gamma_{1.4}^{2/3}\kappa_{10}^{-1/3}\alpha_{-3}^{1/3} f_r^{-2/3}\dot{m}_{-9}^{1/3}\:{\rm cm}.
\end{equation}
Thus the bulk of the pebble population delivered to the dead zone
inner edge will be trapped near the pressure maximum.
The surface density of pebbles will continue to grow. Next we discuss
the implications for planet formation from such a ring.


\subsection{Planet Formation}
We consider two planet formation mechanisms: (1)
gravitational instability; (2) core accretion.

\subsubsection{Via Gravitational Instability}
\label{S:GI}
\begin{figure*}
\begin{center}
\plotone{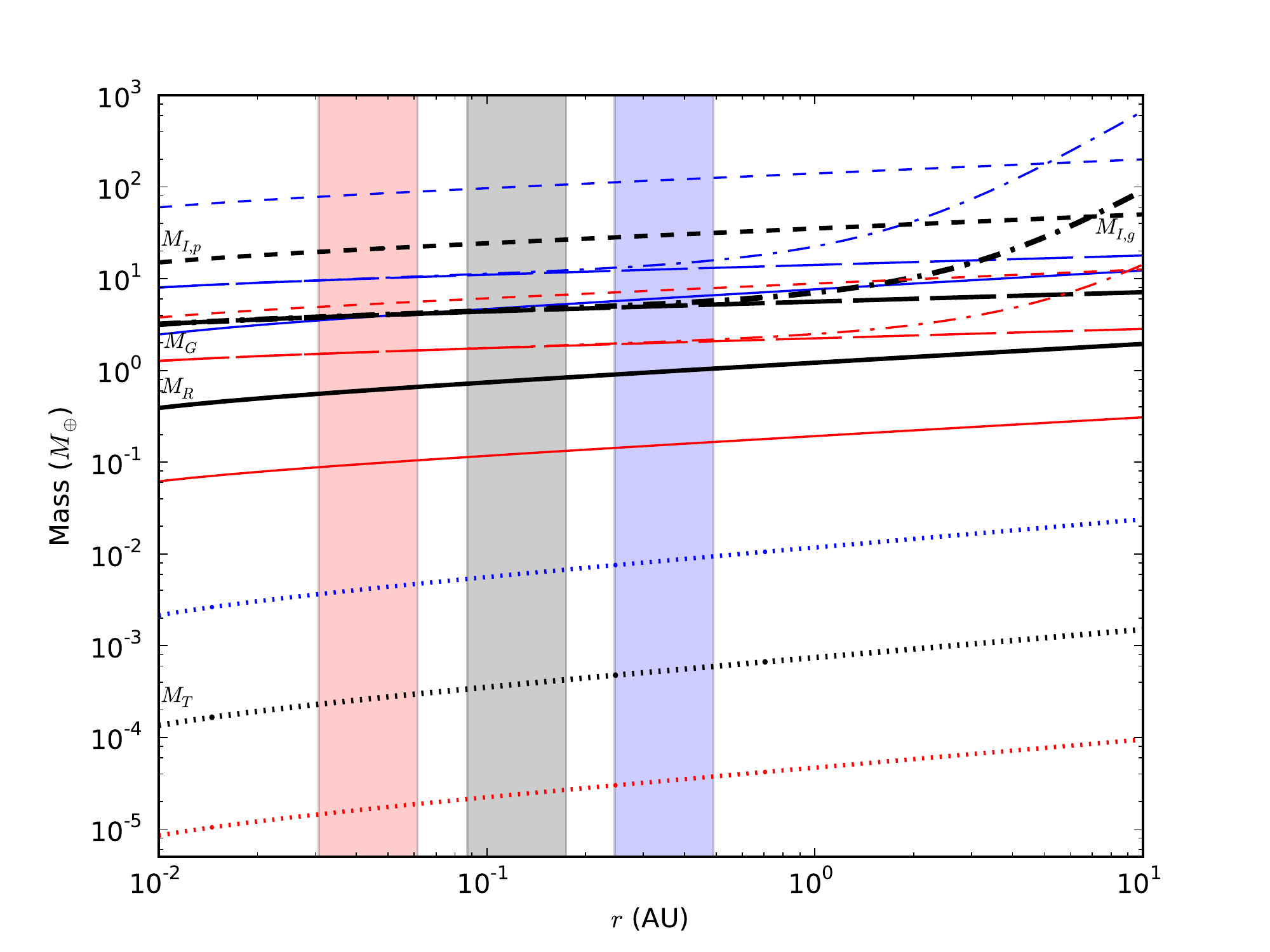}
\caption[$M$ vs $r$]{
Mass scales of planet formation versus distance, $r$, from star for
disks with accretion rate $\dot{m}=10^{-10}, 10^{-9}, 10^{-8}\:\smyr$
(red, black, blue lines, respectively).
Toomre mass $M_T$ (dotted), Toomre ring mass $M_R$ (solid),
gap-opening mass $\Mgap$ (long-dashed), isolation mass in
gas-dominated disk $M_{I, g}$ (dot-dashed), and isolation mass in
pebble-dominated disk $M_{I, p}$ (dashed) are shown. The vertical shaded regions 
indicate the approximate locations for $T=1200$~K, where thermal ionization 
of alkali metals is expected to become important. Red, black, and blue indicate 
$\dot{m}=10^{-10}, 10^{-9}$, and $10^{-8}\:\smyr$, respectively. The right boundaries 
in the shaded regions are at $r_{\rm{1200K}}(\dot{m})$ given by Equation\ \ref{eq:rinner}. 
The left boundaries are at $0.5 r_{\rm{1200K}}(\dot{m})$. The choice of the left boundaries 
is somewhat ad-hoc and indicates that the location for a given temperature can 
be quite uncertain (see text). 
}
\label{fig:theory}
\end{center}
\end{figure*} 

Planets may form via Toomre ring instability of a pebble-dominated
region with $\Sigma(r_0)\simeq \Sigma_p$.  
With reference to the Toomre 
stability parameter for a gaseous disk, 
instability develops
when $Q\equiv\Omega_K\sigma_p/(\pi G\Sigma_p)\lesssim1$, where
$\sigma_p$ is pebble velocity dispersion.  If the mass surface
densities of pebbles is much greater than that of gas, i.e. $\Sigma_p
\gg \Sigma_g$, as is shown below, then their velocity dispersion will
be little affected by any MRI-induced turbulence that is in the
vicinity of the dead zone boundary and 
we expect $\sigma_p<c_s$.

We assume $\sigma_p=\phi_\sigma|v_{r,p}(\taufric=1)|$, with
$\phi_\sigma\sim{\cal O}(1)$, i.e. pebble velocity dispersion is
similar to the maximum drift speed just before delivery to $r_0$. This
inertial limit is expected if there is a sharp decrease in pressure at
the inner dead zone boundary and is thus an upper limit. It is also an
upper limit given that we expect a wide mass spectrum of pebbles to be
delivered, that will have a range of values of $\taufric$.  From
equations~\ref{eq:taufric} and \ref{eq:rinner}, we see
that for fiducial parameters, $\sim$cm-sized pebbles reach the dead
zone inner boundary (if set by thermal ionization) via Epstein drag
with $\taufric\sim 0.01$. Larger, 10-cm-sized pebbles would be in the
Stokes drag regime with Reynolds numbers ${\rm Re} = 2 a_p
v_\Delta/\nu_{\rm mol} < 1$, where $\nu_{\rm mol}$ is the microscopic
(molecular) viscosity $\nu_{\rm mol}\simeq \lambda c_s \simeq c_s
/(n_{\rm H2} \sigma_{\rm H2})$ and $\sigma_{\rm H2}\simeq 2\times
10^{-15}\:{\rm cm^{2}}$. In this case, $C_D = 24 {\rm Re}^{-1}$ and
\begin{eqnarray}
\label{eq:taufric_stokes}
\taufric & \rightarrow & 0.101 a_{p,10}^2 \rho_{p,3} \gamma_{1.4}^{-2/5}\kappa_{10}^{-1/10}\alpha_{-3}^{1/10}\nonumber\\
& \times & m_{*,1}^{7/20}(f_r\dot{m}_{-9})^{-1/5}\rau^{-21/20},
\end{eqnarray}
using
\begin{eqnarray}
\label{eq:Re}
{\rm Re} & \rightarrow & 9.65 a_{p,10} \frac{\taufric}{\taufric+\taufric^{-1}} \gamma_{1.4}^{-4/5}\kappa_{10}^{1/5}\alpha_{-3}^{4/5}\nonumber\\
& \times & m_{*,1}^{1/5}(f_r\dot{m}_{-9})^{3/5}\rau^{-8/5}.
\end{eqnarray}
However, note that at the fiducial location of the dead zone inner
boundary, given by Equation~\ref{eq:rinner}, i.e. 0.178~AU, a
10-cm-sized pebble would have $\taufric=0.62$, implying ${\rm Re} \sim
40$, which is a regime described by a different drag coefficient, $C_D
= 24 {\rm Re}^{-0.6}$ ($1<{\rm Re}<800$), that would predict a value
of $C_D$ about 4 times higher than in the ${\rm Re}<1$ regime.  We see
that an accurate calculation of the mass-averaged value of
$\phi_\sigma$ of the pebbles delivered to a particular dead zone inner
boundary location depends on a model for the size distribution and
thus growth of pebbles. Such a calculation is beyond the scope of the
present paper, and so for simplicity we adopt a fiducial value of $\phi_\sigma =
0.3$, which will characterize the behavior of the population of
pebbles that are being delivered most efficiently by gas drag to the
inner disk.
We note that once pebbles reach the location of the pressure maximum,
$r_0$, there are other processes such as continued gas drag and
pebble-pebble collisions that can also act to reduce $\sigma_p$.  On
the other hand, excitation of velocity dispersion by interaction with
any turbulence present in the gas would tend to increase
$\phi_\sigma$.

We thus express the mass surface density of pebbles at the time of development
of gravitational instability as
\begin{eqnarray}
\label{eq:Sigmap1}
\Sigma_p&=&\frac{\phi_\sigma|v_{r,p}|\Omega_K}{\pi GQ}=\frac{\phi_\sigma k_Pc_s^2}{2\pi GQr}\nonumber\\
&=&\frac{3^{1/5}}{4\pi^{7/5}}\phi_\sigma k_PQ^{-1}
\left(\frac{\mu}{\gamma k_B}\right)^{-4/5}
\left(\frac{\kappa}{\sigmasb}\right)^{1/5}
\alpha^{-1/5}\nonumber\\
&\times&G^{-7/10}m_*^{3/10}
\left(f_r\dot{m}\right)^{2/5}
r^{-19/10}\\
&\rightarrow&1.54 \times 10^3 \phi_{\sigma,0.3} Q^{-1}\gamma_{1.4}^{4/5}\kappa_{10}^{1/5}\alpha_{-3}^{-1/5}\nonumber\\
 & \times & m_{*,1}^{3/10}(f_r\dot{m}_{-9})^{2/5}\rau^{-19/10}\:\gcc.\nonumber
\end{eqnarray}
The above mass surface density is much higher than that of the gas,
\begin{eqnarray}\label{eq:Sigmagas}
\Sigma_g&=&\frac{2}{3^{6/5}\pi^{3/5}}
\left(\frac{\mu}{\gamma k_B}\right)^{4/5}
\left(\frac{\kappa}{\sigmasb}\right)^{-1/5}
\alpha^{-4/5}\nonumber\\
&\times&(Gm_*)^{1/5} 
\left(f_r\dot{m}\right)^{3/5}
r^{-3/5}\\
&\rightarrow&106\gamma_{1.4}^{-4/5}\kappa_{10}^{-1/5}\alpha_{-3}^{-4/5}m_{*,1}^{1/5}(f_r\dot{m}_{-9})^{3/5}\rau^{-3/5}\:\gcc,\nonumber
\end{eqnarray}
with ratio of $\Sigma_p/\Sigma_g$ given by:
\begin{eqnarray}\label{eq:Sigmaratio}
\frac{\Sigma_p}{\Sigma_g}&=&\frac{3^{7/5}}{8\pi^{4/5}}\frac{k_P\phi_\sigma}{Q}\left(\frac{\mu}{\gamma k_B}\right)^{-8/5}\left(\frac{\kappa}{\sigmasb}\right)^{2/5}\alpha^{3/5}\nonumber\\
&\times&G^{-9/10}m_*^{1/10}\left(f_r\dot{m}\right)^{-1/5}r^{-13/10}\\
&\rightarrow&14.5\phi_{\sigma,0.3}\gamma_{1.4}^{8/5}\kappa_{10}^{2/5}\alpha_{-3}^{3/5}m_{*,1}^{1/10}(f_r\dot{m}_{-9})^{-1/5}\rau^{-13/10}.\nonumber
\end{eqnarray}
Thus at the time of the development of gravitational instability we do
not expect the pebble velocity dispersion to be significantly
influenced by that of the gas.

The most unstable radial length scale in the pebble ring is
$\lambda_{T}=2\sigma_p^2/(G\Sigma_p)$.  An approximate estimate for
the minimum mass associated with this scale is the Toomre mass
\begin{eqnarray}
\label{eq:mt}
M_{T}&\equiv&\Sigma_p\lambda_{T}^2=\frac{\pi\phi_\sigma^3k_P^3Qc_s^6r^3}{2G^3m_*^2}\nonumber\\
&=&\frac{3^{3/5}}{2^{4}\pi^{1/5}}
\phi_\sigma^3k_P^3Q
\left(\frac{\mu}{\gamma k_B}\right)^{-12/5} 
\left(\frac{\kappa}{\sigmasb}\right)^{3/5}
\alpha^{-3/5}\nonumber\\
&\times&G^{-21/10}m_*^{-11/10}
\left(f_r\dot{m}\right)^{6/5}
r^{3/10}\\
&\rightarrow&7.60\times 10^{-4}\phi_{\sigma,0.3}^3Q\gamma_{1.4}^{12/5}\kappa_{10}^{3/5}\alpha_{-3}^{-3/5}\nonumber\\
 & \times & m_{*,1}^{-11/10}(f_r\dot{m}_{-9})^{6/5}\rau^{3/10}\:M_\oplus.\nonumber
\end{eqnarray}
However, this is likely to be a lower limit on the mass accumulated by gravitational instability. 
Note that the orbital timescale $t_{\rm{orb}}$ for the $M_T$-mass bodies 
at $r_0$ is much shorter than $t_{\rm{drift}}$. Hence, 
we expect that 
the bulk of the ring
material will be gathered into a single planet
with Toomre ``ring mass''
\begin{eqnarray}
\label{eq:mr}
M_R&\equiv&2\pi r\lambda_T\Sigma_p=4\pi r\phi_\sigma^2\frac{|v_{r,p}|^2}{G}=\frac{\pi\phi_\sigma^2k_P^2r^2c_s^4}{G^2m_*}\nonumber\\
&=&\frac{3^{2/5}\pi^{1/5}}{2^{2}}
\phi_\sigma^2k_P^2
\left(\frac{\mu}{\gamma k_B}\right)^{-8/5}
\left(\frac{\kappa}{\sigmasb}\right)^{2/5}
\alpha^{-2/5}\nonumber\\
&\times&G^{-7/5}m_*^{-2/5}
\left(f_r\dot{m}\right)^{4/5}
r^{1/5}\\
&\rightarrow&1.23\phi_{\sigma,0.3}^2\gamma_{1.4}^{8/5}\kappa_{10}^{2/5}\alpha_{-3}^{-2/5}m_{*,1}^{-2/5}(f_r\dot{m}_{-9})^{4/5}\rau^{1/5}\:M_\oplus.\nonumber
\end{eqnarray}
Figure~\ref{fig:theory} shows $M_T(r)$ and $M_R(r)$ in disks around a
solar-mass star with $\dot{m}=10^{-10},10^{-9}$ and $10^{-8}\:\smyr$. 

The ratio of the ring mass to the Toomre mass is
\begin{eqnarray}
\label{eq:mrtomtoomre}
\frac{M_R}{M_T}&\equiv&\frac{4\pi^{2/5}}{3^{1/5}}
(\phi_\sigma k_P Q)^{-1}
\left(\frac{\mu}{\gamma k_B}\right)^{4/5}
\left(\frac{\kappa}{\sigmasb}\right)^{-1/5}
\alpha^{1/5}\nonumber\\
&\times&(Gm_*)^{7/10}
\left(f_r\dot{m}\right)^{-2/5}
r^{-1/10}\\
&\rightarrow&1620\phi_{\sigma,0.3}^{-1}Q^{-1}\gamma_{1.4}^{-4/5}\kappa_{10}^{-1/5}\alpha_{-3}^{1/5}\nonumber\\
 & \times &m_{*,1}^{7/10}(f_r\dot{m}_{-9})^{-2/5}\rau^{-1/10}.\nonumber
\end{eqnarray}
Thus growth to $M_R$ from $M_T$ involves an increase in mass by a
large factor. Detailed investigation of this stage requires numerical
simulation, but should fall within two limits: (1) Pebble accretion by
the first Toomre mass protoplanet; (2) Oligarchic growth from a
population of many Toomre mass protoplanets born together at the same
orbital radius. In the first case, a ring mass planet is formed on a
near circular orbit from the pebble ring. In the second case, for
these relatively low mass protoplanets that are quite close to the central
star, strong encounters often lead to physical collisions and strong
scattering is not likely. 
Moreover, the observed low dispersion of orbital inclination angles
in the STIPs planets would require this limit to apply.

The planet may continue to grow beyond $M_R$ since it is still
embedded in a gaseous disk that is also still delivering pebbles.  As
discussed below (\S\ref{S:subsequent}), truncating such accretion may
require either the planet becoming massive enough to open a gap, or
migrating away from the pressure maximum via inward Type~I migration
\citep[e.g.,][]{2004ApJ...606..520M}.

\subsubsection{Via Core Accretion}

Alternatively, planets may form via core accretion from the rich
supply of solids in the pebble ring. However, because of difficulties
in sticking meter-sized pebbles together, the first step of forming
planetesimals likely requires larger-scale streaming instabilities
\citep[e.g.,][]{2005ApJ...620..459Y} or gathering of material in
vortices \citep[e.g.,][]{2006A&A...446L..13V}. Collisional runaway
growth of a protoplanet may then occur from this planetesimal
population. The practical difference between this formation scenario
and that involving gravitational instability is that the minimum planet mass is now
$\ll\:M_T$.  However, since fiducial values of $M_T\ll\:M_\oplus$, it
is difficult to distinguish these scenarios observationally.

\subsection{Migration, Gap Opening, Dead Zone Retreat and Subsequent Planet Formation}\label{S:subsequent}

Once a planet has formed from the pebble ring, we envisage two
potential subsequent evolutionary scenarios: (1) Efficient Type I
migration of the planet into the MRI-active region (\S\ref{S:type1}),
followed by formation of another pebble ring and eventually another
planet at the dead zone inner boundary, at approximately fixed
location in the disk; (2) Inefficient Type I migration and/or rapid
growth of the planet to a mass capable of opening a gap in the disk
(\S\ref{S:gap}), followed by dead zone retreat and formation of a new
pebble ring and planet further out in the disk. As discussed later in
\S\ref{S:subsequent}, the global reservoir of pebbles places
constraints on both of these scenarios.

\subsubsection{Birth and Migration from a Fixed Parent Pebble Ring}
\label{S:type1}

If $M_R$ is smaller than the mass needed to open a gap in the disk,
$M_G$ (see \S\ref{S:gap}), 
then such a planet will undergo Type I migration
\citep{1997Icar..126..261W}. Detailed analysis of the ultimate fate of
a planet undergoing Type I migration is an active area of research
\citep[e.g.,][]{2010MNRAS.401.1950P}.
The rate and even the direction of migration in the region near the
dead-zone-MRI-active-zone boundary will depend on the details of the
complicated local density and temperature profiles, which in turn
depend on the changes in $\alpha$ and $\kappa$. Thus 
here we simply discuss the expected qualitative behavior of these
migrating planets and the implications for such migration on
observable planet properties.

For our fiducial disk model ($\Sigma \sim r^{-3/5}$;
Equation~\ref{eq:Sigmagas}) the co-orbital torque is positive, i.e.,
the co-orbital torques would result in an outward migration (for a
review see \citealt{2011exop.book..347L}). If the co-orbital
torques are saturated, torque due to the Lindblad resonances (LR)
dominate and result in inward migration of the planet. However, due to
the steep $r$-dependence of the angular momentum of a planet's orbit,
the migration rate $\dot{r}_{\rm{Type\ I}}$ decreases with decreasing
$r$ for typical disk density profiles.
For example, for our fiducial disk model the Type\ I migration rate
due to the LRs for a given planet mass $M_{\rm{pl}} \sim M_R$ is
$\dot{r}_{\rm{Type\ I}} \sim r^{9/10}$ if $h/R_H<1$, and $\sim
r^{4/5}$ if $h/R_H>1$, where $h$ is the disk scale height, and
$R_H\equiv (M_{\rm pl}/[3m_*])^{1/3}r$ is the Hill sphere of the
migrating planet \citep{2011exop.book..347L}. As the planet migrates
into the MRI active region by crossing $r_0$, the dead-zone inner
boundary, it finds itself in a disk with much lower $\Sigma_g \sim
\alpha^{-4/5}$ (Equation~\ref{eq:Sigmagas}) because of the
potentially orders of magnitude larger $\alpha$ in the MRI-active
region compared to $\alpha$ inside the dead-zone.  The low-$\Sigma_g$
also would result in a low $\dot{r}_{\rm Type\ I} \sim \Sigma_g$.  The
temperature gradient both from the $r$-dependence and change in
$\kappa$ at the boundary (Equation~\ref{eq:temp}) can also contribute
to an outward net torque component on the planet
\citep[e.g.,][]{2006A&A...459L..17P,2010MNRAS.401.1950P}.

If there is efficient inward Type\ I migration of the planet away from
the dead zone inner boundary, then conditions may be set up for
re-forming a pebble ring at the associated pressure maximum. A whole
series of planets may form sequentially at $r_0$, which then migrate
inwards to form a compact planetary system. If the disk properties
($\dot{m}$, $r_0$) are relatively steady, then the resulting planetary
masses, compositions and densities may also be quite similar.

The ultimate change in the planetary orbits due to Type\ I migration
will depend on both the rate of migration and the amount of time
available for this process.  Once inside the MRI-active region,
further growth of these planets via accretion appears to be difficult,
since (1) pebbles remain trapped at the dead zone inner boundary; (2)
the gas is hot ($\gtrsim 1200$~K) and hence harder to accrete.
Type I migration may be limited by the time needed to form
a planet (at $r_0$ or further out in the dead zone) that is massive
enough to open a gap in the gas disk and thus lead to starvation of
the inner gas disk and its depletion via viscous clearing. Strong
stellar magnetic fields may truncate the gas disk at a few stellar
radii, i.e. $\sim 10R_\odot \sim 0.05$~AU, and this could set an inner
limit for Type\ I migration.

The main prediction of the strong migration scenario is the presence
of planets at locations inside the inner dead zone boundary, although
this location is uncertain (\S\ref{S:pmax}) and depends on the disk
accretion rate. The implications of the observed KPCs for this
scenario are discussed in \S\ref{S:Kepler}. Next we consider the case
of weak migration coupled with efficient growth of planets leading to
gap opening.

\subsubsection{Gap Opening and Dead Zone Retreat}
\label{S:gap}
The process of gap opening by a planet involves it clearing a region
over which it has a dominant gravitational influence compared to the
star. A planet of mass $\Mpl$ orbiting in a disk has strong
gravitational influence on orbits with impact parameters falling
approximately within its Hill sphere, $R_H$:
\clearpage
\begin{eqnarray}
\label{eq:rhill}
\frac{R_H}{r}&\equiv&\left(\frac{\Mpl}{3m_*}\right)^{1/3} = \left(\frac{M_R}{3m_*}\right)^{1/3} \nonumber\\
&\rightarrow&0.0107\phi_{\sigma,0.3}^{2/3}\gamma_{1.4}^{8/15}\kappa_{10}^{2/15}
\alpha_{-3}^{-2/15}m_{*,1}^{-7/15}\nonumber\\
&\times&(f_r\dot{m}_{-9})^{4/15}\rau^{1/15}.
\end{eqnarray}
We assume the planet accretes material out to impact parameter
$\phi_HR_H$, where $\phi_H\sim3$
\citep[e.g.,][]{1987Icar...69..249L,1998Icar..131..171K}. 
The fractional width of the Toomre unstable ring, $\lambda_T/r$, is given by
\begin{eqnarray}
\frac{\lambda_T}{r}&=&2\pi Q \sigma_p / v_K \nonumber\\
& = & \frac{3^{1/5}\pi^{3/5}}{2}\phi_\sigma k_PQ
\left(\frac{\mu}{\gamma k_B}\right)^{-4/5}
\left(\frac{\kappa}{\sigmasb}\right)^{1/5}
\alpha^{-1/5}\nonumber\\
&\times&(Gm_*)^{-7/10}
\left(f_r\dot{m}\right)^{2/5}
r^{1/10}\\
&\rightarrow&3.41\times 10^{-3}\phi_{\sigma,0.3} Q\gamma_{1.4}^{4/5}\kappa_{10}^{1/5}\alpha_{-3}^{-1/5}m_{*,1}^{-7/10}\nonumber\\
& \times & (f_r\dot{m}_{-9})^{2/5}\rau^{1/10},\nonumber
\end{eqnarray}
which is about a factor of three smaller than $R_H/r$ for all relevant $r$ for
our fiducial disk ($R_H/\lambda_T\propto\dot{m}^{-2/15}r^{-1/30}$).
Thus after the $M_R$-mass planet forms from the ring, it will still dominate 
regions of the disk beyond the initial ring width, and we expect the planet's mass to 
grow beyond $M_R$. 

We estimate final isolation mass in two ways. First, we evaluate
the isolation mass in a pebble-rich disk, $M_{I,p}$, as $M_R$
plus additional accreted mass from sweeping-up a disk with
$\Sigma\simeq\Sigma_p$ over impact parameters out to $\phi_HR_H$. This
case is relevant if the annular width of the region that had $\Sigma$
enhanced by pebble drift is $\gtrsim\phi_HR_H$.
In this case
\begin{eqnarray}
\label{eq:MIa}
M_{I,p}/M_R&=&1+\phi_{H,p}R_H(\Mpl=M_{I,p})/\lambda_T\nonumber\\
&&\simeq\phi_{H,p}R_H(\Mpl=M_{I,p})/\lambda_T,
\end{eqnarray}
implying
\begin{eqnarray}\label{eq:MI}
M_{I,p}&=&\frac{1}{2^{3/2}3^{1/5}\pi^{3/5}}\left(\frac{\phi_{H,p}\phi_\sigma k_P}{Q}\right)^{3/2}
\left(\frac{\mu}{\gamma k_B}\right)^{-6/5}\\
&\times& \left(\frac{\kappa}{\sigmasb}\right)^{3/10}\alpha^{-3/10}
G^{-21/20}m_*^{-1/20}
\left(f_r\dot{m}\right)^{3/5}
r^{3/20}\nonumber\\
&\rightarrow&35.7\left(\frac{\phi_{H,p,3}\phi_{\sigma,0.3}}{Q}\right)^{3/2}\gamma_{1.4}^{6/5}\kappa_{10}^{3/10}\alpha_{-3}^{-3/10}m_{*,1}^{-1/20}\nonumber\\
&\times&(f_r\dot{m}_{-9})^{3/5}\rau^{3/20}\:M_\oplus,\nonumber
\end{eqnarray}
where $\phi_{H,p,3}\equiv\phi_{H,p}/3$. The approximation assuming
$M_{I,p}\gg\:M_R$ in Equation\ \ref{eq:MIa} is thus verified. $M_{I,p}$ is also shown in
Fig.~\ref{fig:theory}. Note that, although these can be close to Jovian-mass
planets, they would have approximately terrestrial compositions. As
shown below, this mass would also be sufficient to open an isolating
gap with the gas disk.

Second, if the width of the pebble-enhanced ($\Sigma=\Sigma_p$)
annulus is $\ll\phi_HR_H$, then the isolation mass, $M_{I,g}$, is set
by accretion from a gas-dominated disk. The planet needs to first
reach mass, $\Mgap$, sufficient to open a gas gap.
We estimate this via the viscous-thermal criterion
\citep{1993prpl.conf..749L},
\begin{eqnarray}\label{eq:Mgap}
M_{G}&=&\frac{\phi_{G}40{\nu}m_*}{r^2\Omega_K}\nonumber\\
&=&20\frac{3^{1/5}}{\pi^{2/5}}\phi_{G}\left(\frac{\mu}{\gamma k_B}\right)^{-4/5}\left(\frac{\kappa}{\sigmasb}\right)^{1/5}\nonumber\\
&\times&\alpha^{4/5}G^{-7/10}m_*^{3/10}\left(f_r\dot{m}\right)^{2/5}r^{1/10}\\
&\rightarrow&5.67\phi_{G,0.3}\gamma_{1.4}^{4/5}\kappa_{10}^{1/5}\alpha_{-3}^{4/5}m_{*,1}^{3/10}(f_r\dot{m}_{-9})^{2/5}\rau^{1/10}\:M_\oplus,\nonumber
\end{eqnarray}
where we adopt $\phi_{G}=0.3$ based on simulations of
\citep{2013ApJ...768..143Z}, who also find $\phi_{G}$ depends on net
vertical disk B-field strength.  The ratio $M_R/M_G$ is given by
\begin{eqnarray}\label{eq:MRtoMgap}
\frac{M_R}{M_G}&=&\frac{3^{1/5}\pi^{3/5}}{80} \frac{\phi_\sigma^2 k_P^2}{\phi_{G}} \left(\frac{\mu}{\gamma k_B}\right)^{-4/5}\left(\frac{\kappa}{\sigmasb}\right)^{1/5}\nonumber\\
&\times&\alpha^{-6/5}(Gm_*)^{-7/10}\left(f_r\dot{m}\right)^{2/5}r^{1/10}\\
&\rightarrow&0.217 \phi_{G,0.3}^{-1}\phi_{\sigma,0.3}^2 k_P^2 \gamma_{1.4}^{4/5}\kappa_{10}^{1/5}\alpha_{-3}^{-6/5}m_{*,1}^{-7/10}\nonumber\\
 & \times & (f_r\dot{m}_{-9})^{2/5}\rau^{1/10}\:M_\oplus.\nonumber
\end{eqnarray}
Note that both $M_G$ and $M_R/M_G$ are quite sensitive to the value of $\alpha$, 
which is quite uncertain, especially at the location relevant for pebble
  ring formation near the dead zone boundary.
Nevertheless, for our fiducial disk $M_{R}\lesssim M_G$, so we expect
that some additional accretion would be needed after formation from
the ring mass before a gap could be opened. This phase would allow an
opportunity for Type I migration (\S\ref{S:type1}). If gas is able to
cool and join the planet it could also lead to accretion of both gas
and pebbles, thus leading to lower density planets. However, given the
uncertainties in parameters, such as $\alpha$, we can also imagine
situations where $M_R\gtrsim M_G$, and a gap would be opened
simultaneously with planet formation from the pebble ring.

The gaps seen in the simulations of \citet{2013ApJ...768..143Z} (i.e. for $\phi_G\simeq
0.3$) are relatively shallow (deeper gaps will be achieved with larger
values of $\phi_G$), but still this may be sufficient to allow
additional penetration of X-rays that may increase the ionization
fraction to activate the MRI and thus cause the pressure maximum
associated with the dead zone boundary to move outwards. Such a
scenario, discussed below, could lead to a truncation in the supply of
pebbles to the planet. This supply may also be impeded by the pressure
maximum associated with the outer edge of the gap \citep[e.g.,][]{2007ApJ...660.1609M}.

In the process of opening a deep, well-cleared gap that isolates the
planet from further accretion, the planet will likely accrete an
additional gas mass by sweeping-up an annulus of a few ($\phi_{H,g}$)
Hill radii,
\begin{eqnarray}
dM_{g}& = &2\pi{r}\phi_{H,g}R_H\Sigma_g\nonumber\\
&=&2\pi r^2\phi_{H,g}(M_{\rm pl}/(3m_*))^{1/3}\Sigma_g,
\end{eqnarray}
where $\Sigma_g$ is given by Equation~\ref{eq:Sigmagas}.
The final isolation mass of the planet in a gas disk is thus
\begin{equation}
M_{I,g} = {\rm max}(M_R,\Mgap)+ dM_g.
\end{equation}
The solution of the above equation for $M_{I,g}$ for $\phi_{H,g}=3$ is shown in
Fig.~\ref{fig:theory}. For $\rau\lesssim1$ there is only a very minor
enhancement in mass beyond $M_G$.
For $\rau\gtrsim3$, the planet gains
most of its eventual mass in these final stages of opening a gas gap.
These considerations suggest that $M_R/M_{I,g}$ declines with radius,
so that outer planets will tend to be of lower density.


Once the first planet has formed and opened a gap, we expect that
interior disk material, which is mostly in an MRI-active region, will
rapidly accrete on a local viscous time. The dead zone boundary should
then retreat outwards, since protostellar X-rays will now be able to
penetrate further. The same processes that formed the first planet,
i.e. collection of pebbles at a pressure maximum, should then operate
to form a second planet, assuming there is still a supply of pebbles
from the outer disk.

The pressure maximum associated with the first planet's outer gap edge
sets a minimum separation of the location of the next planet to be
$\sim\phi_HR_H$. 
However, if the MRI-active, inner disk makes a transition to being
completely cleared, 
then we expect a large reduction in the absorbing
column and thus perhaps a large shift in the location of the dead zone
inner boundary, especially relative to $\phi_H R_H$, since $\phi_H R_H
/ r \ll 1$.
An accurate estimate of the distance of dead zone retreat would
involve a sophisticated calculation of the ionization, thermal and
magnetic field structure of the disk, as a gap and inner hole are
established. 
We defer such a calculation to a future paper and for the moment
simply assume a new dead zone inner boundary pressure maximum will be
established at least $\phi_HR_H$ from the first planet 
but likely significantly further. 

Assuming a steady disk accretion rate and constant value of $\alpha$,
the masses of planets forming from an initially gravitationally
unstable ring should follow the radial dependencies of
Equation~\ref{eq:Mgap} ($M_{I,g}\simeq\:M_G\propto\:r^{1/10}$ for
$r\lesssim1$~AU) for isolation in a gas-dominated disk where
$M_G>M_R$. If $M_R>M_G$, then the mass scaling with radius would be
expected to follow Equation~\ref{eq:mr}
($M_{I,g}\simeq\:M_R\propto\:r^{1/5}$ for $r\lesssim1$~AU).  In a
pebble-dominated disk, then the the masses would be described by
Equation~\ref{eq:MI} ($M_{I, p}\propto\:r^{3/20}$): these masses
tend to always be enough to open a gap. 
These are all similar, relatively flat scalings with the orbital radius. These
dependencies can be tested against observed planetary systems
(\S\ref{S:Kepler}), but with the caveat that there is the possibility
of the efficient Type I migration scenario (\S\ref{S:type1}), in which
planetary orbits are shrunk from the location of the parent pebble
ring.

\subsubsection{Constraints from the Global Disk Pebble Reservoir}

In either limit of efficient or inefficient Type\ I migration,
subsequent planet formation 
requires continued pebble drift to $r_0$, which will be reduced once
the reservoir of disk solids is depleted. The mass in solids initially
contained in the gas disk within radius $r_1\gg\:r_0$ is
\begin{eqnarray}
M_s(<r_1)&=&\int^{r_1}f_s2\pi r\Sigma_gdr\nonumber\\
&=&\frac{20\pi^{2/5}}{3^{6/5}7}f_s
\left(\frac{\mu}{\gamma k_B}\right)^{4/5}
\left(\frac{\kappa}{\sigmasb}\right)^{-1/5}
\alpha^{-4/5}\nonumber\\
&\times&(Gm_*)^{1/5}
\dot{m}^{3/5}
r_1^{7/5}\\
 & \rightarrow & 0.178 f_{s, -2} \gamma_{1.4}^{-4/5} \kappa_{10}^{1/5} \alpha_{-3}^{-4/5} m_{*,1}^{1/5} \dot{m}_{-9}^{3/5}  r_{\rm{1,AU}}^{7/5}\:M_\earth.\nonumber
\end{eqnarray}
Assuming the first planet forms with mass $M_R=\epsilon_pM_s(<r_1)$
with efficiency $\epsilon_p=0.5$, we estimate the radius $r_1$ that 
becomes depleted of pebbles:
\begin{eqnarray}\label{eq:rone}
r_1&=&\left(\frac{7}{5}\right)^{5/7}\frac{3^{8/7}}{2^{20/7}\pi^{1/7}}\frac{(\phi_\sigma k_P)^{10/7}}{(f_s\epsilon_p)^{5/7}}
\left(\frac{\mu}{\gamma k_B}\right)^{-12/7}
\left(\frac{\kappa}{\sigmasb}\right)^{3/7}\nonumber\\
&\times&\alpha^{2/7}
G^{-8/7}m_*^{-3/7}
\dot{m}^{1/7}
r_0^{1/7}\\
&\rightarrow&6.55\frac{\phi_{\sigma,0.3}^{10/7}\gamma_{1.4}^{12/7} \kappa_{10}^{3/7}}{(f_{s,0.01}\epsilon_{p,0.5})^{5/7}}\alpha_{-3}^{2/7}m_{*,1}^{-3/7}\dot{m}_{-9}^{1/7}r_{\rm{0,AU}}^{1/7}\:\rm{AU}.\nonumber
\end{eqnarray}
Note we have adopted a single value of $\alpha$ for the disk out to
$r_1$. If the dead zone outer boundary has a radial extent $<r_1$,
then this estimate would need to be modified, leading to an increased
value of $\alpha$ in the outer region and thus a larger value of
$r_1$. Equation \ref{eq:rone} shows that a fairly large region of
the disk is needed to supply the mass of pebbles to form a Toomre ring
mass planet, comparable to the outer scales predicted for dead zones
\citep[e.g.,][]{2013ApJ...764...65M,2013ApJ...765..114D}.  
Formation of a 
series of super-Earth mass planets from pebbles could require initial
protoplanetary disks extending to $\sim 100$~AU.

Pebble drift can also be reduced if an outer planet forms, e.g., via
regular core accretion, gaseous gravitational instability, or
gravitational instability of an outer pebble ring captured in a local
pressure maximum. If massive enough, such a planet would interrupt the
supply of pebbles, i.e., they would be depleted from the disk interior
to this planet. However, we expect regular core accretion in the outer
disk to be slower than pebble drift to the inner region, and indeed
inhibited by pebble drift. Gaseous gravitational instability is
unlikely to operate within $\sim$10--100~AU unless disks are very
massive \citep[e.g.,][]{2005ApJ...621L..69R}. If an outer pebble ring
forms first before an inner ring is established at $r_0$, then that
process can be viewed as a scaled-up version of the theory presented
here. Outer pebble ring formation may be induced by pressure maxima
induced by sudden opacity changes \citep[e.g. at ice lines]{dwd13,bf13} 
or MRI activity changes \citep[e.g., due to gas-phase metal freeze out;][]{2013ApJ...765..114D}. 
The relative efficiency of inner versus outer pebble ring formation
may depend sensitively on disk properties, including $\dot{m}$ and
initial magnetization, leading to distinct classes of planetary
systems, e.g. STIPs versus Solar-System analogs.

Once inside-out planet formation via pebble rings finishes, much of
the remaining gas in the disk will be likely accreted by the outermost
planet, eventually crossing its gap
\citep[e.g.,][]{2013ApJ...769...97U}, to form a gas giant, which would then
deviate from the above analytic $\Mpl-r$ relations.

\begin{figure*}
\begin{center}
\plotone{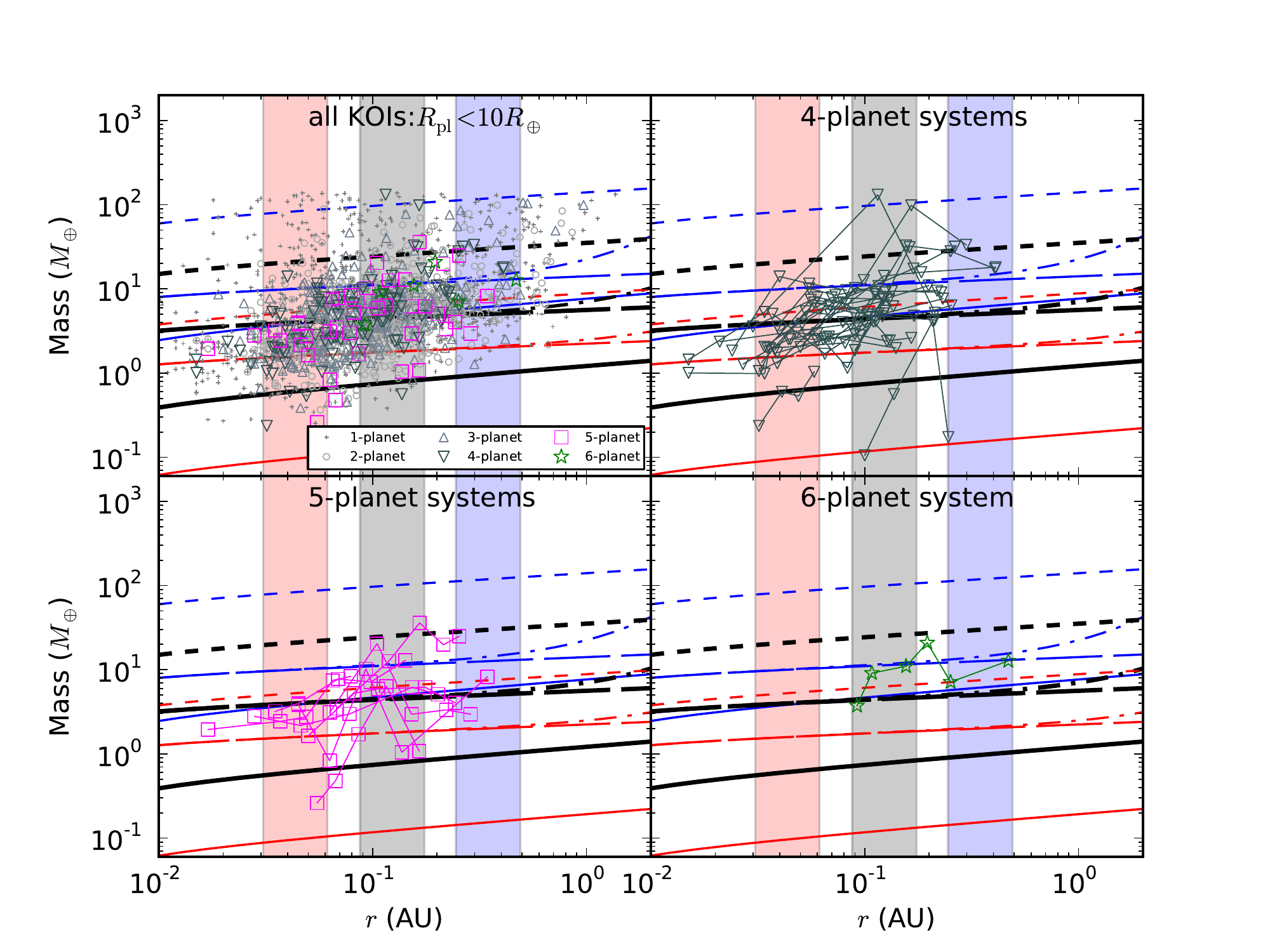}
\caption[$M$ vs $r$ with KPCs]{
Lines and shaded regions have the same meaning as in Figure \ \ref{fig:theory}, 
but zoomed to a narrower mass range. (a) Top-left:
KPCs with $\Rpl<10\:R_\oplus$ are shown
from \citet{2013ApJS..204...24B} (16-month data release).
(b)~Top-right: Only 4-planet systems are shown. (c)~Bottom-left: Only
5-planet systems are shown. (d)~Bottom-right: Only the 6-planet system
is shown. Note, here the KPC masses are approximate estimates using a simple scaling-law 
with radius (see text).
}
\label{fig:KOI}
\end{center}
\end{figure*} 
%

\section{Comparison to Kepler Systems}\label{S:Kepler}

\begin{figure*}
\begin{center}
\plotone{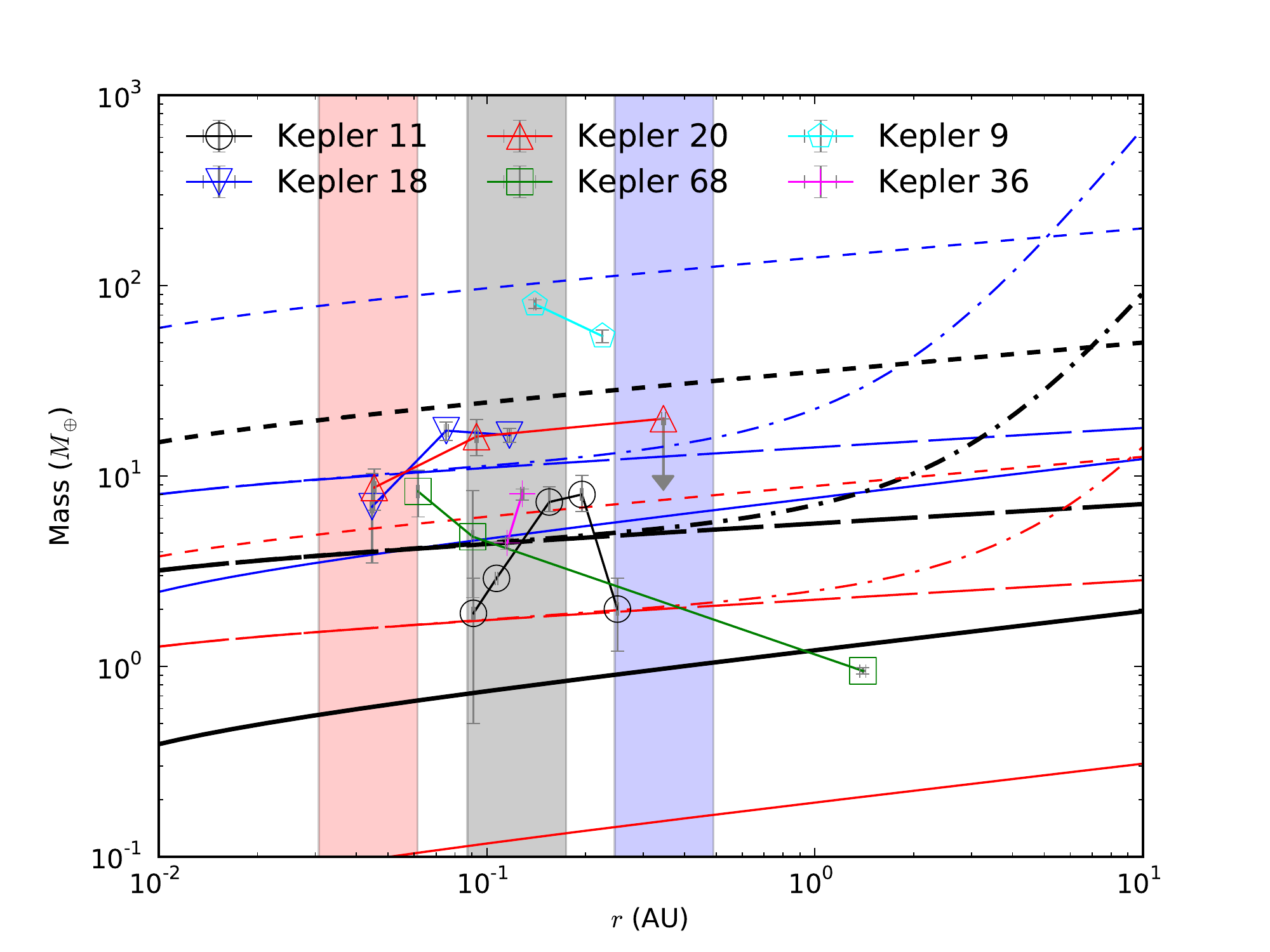}
\caption[$M$ vs $r$]{
Similar to Figure\ \ref{fig:KOI}, but showing the 6 \kepler\ systems
with direct mass measurements. 
}
\label{fig:TTV}
\end{center}
\end{figure*} 

Figure~\ref{fig:KOI}a shows $M_R$, $M_G$, $M_{I,g}$, and $M_{I,p}$ for
$\dot{m}= 10^{-10}, 10^{-9}$ and $10^{-8}\:\smyr$ together with the
KPCs, whose masses are crudely estimated using a power-law
$\Mpl=M_\oplus(R_{\rm pl}/R_\oplus)^{2.06}$
\citep{2011ApJS..197....8L}.  Focusing on STIPs, we discard planets
with $\Rpl\geq10\:R_\oplus$ (none are in multi-transiting systems).
The estimated KPC masses are similar to those expected from the
fiducial model of inside-out planet formation.  However, since
$M_{I,g}\simeq M_R\propto\dot{m}^{4/5}$ (Equation~\ref{eq:mr}) or
$M_{I,g}\simeq M_G\propto\dot{m}^{2/5}$ (Equation~\ref{eq:Mgap}) and
$M_{I,p}\propto\dot{m}^{3/5}$ (Equation~\ref{eq:MI}), a range in masses
could occur at a given $r$ if $\dot{m}$ varies. Such variation is
expected from system to system and even over time within a given
system during planet formation.

Radial dependence of relative planetary masses in a given system
provides a more powerful test, since this removes some systematic
uncertainties resulting from system to system variation, such as $m_*$
and perhaps some dispersion in $\dot{m}$. The twenty-eight 4-planet systems, the
eight 5-planet systems and the single 6-planet system are shown in
Figs.~\ref{fig:KOI}~b, c, and d, respectively.  Fitting a power-law
$\Mpl\propto r^{k_M}$ to these individual systems, we find
$k_M=0.92\pm0.63,0.78\pm0.64,0.50$ for the 4, 5, 6-planet systems
(uncertainty reflects sample dispersion), respectively. These results
are consistent with the theoretical predictions, with caveats that
there may be large systematic errors in these mass estimates and
current orbits may differ from formation orbits due to migration.

Some KPCs are observed interior to the estimated dead-zone boundaries
in our fiducial disk model (Figure\ \ref{fig:KOI}), although these
locations are quite uncertain.
This would imply that some degree of migration has occurred, such as
described in \S\ref{S:type1} or after gap opening, via Type II migration.

A subset of the KPCs have directly measured masses, primarily by
transit timing variations 
\citep[TTV; e.g.,][]{2006ApJ...646..505B,2010Sci...330...51H,2011ApJS..197....7C,2012Sci...337..556C,2012ApJ...749...15G,2013arXiv1303.0227L}.
Figure~\ref{fig:TTV} shows the theoretical $\Mpl-r$ relations along
with these systems (see also Table~\ref{tab:list}).  Averaging these 6
systems, $k_M=1.0\pm2.1$. Averaging all adjacent pairs,
$k_M=0.47\pm2.7$. These values are consistent with scalings for
$M_{I,g}\simeq M_G$ ($k_M=0.1$ for $r\lesssim1$~AU) or $M_{I,p}$
($k_M=0.15$), but more data are required for a more stringent test.
There is a real and significant dispersion in the values of $k_M$ seen
in adjacent planetary pairs within the systems with $\geq3$ planets,
which, in the context of inside-out planet formation, would require
variation of $\dot{m}$ of factors of a few during formation of the
system.

Planetary densities show wide dispersion, but a tendency to decrease
with $r$ (Table~\ref{tab:list}). Some relatively low densities are
seen, which would require $M_{\rm pl}\gg M_R$ and imply that gas
accretion could occur onto the initial core. Even for higher density
systems, models of rocky cores surrounded by residual H/He atmospheres
are needed for comparison of the theory with these data. Evolution due
to atmospheric evaporation may also complicate such comparisons
\citep{2013arXiv1303.3899O}. 

\begin{figure*}
\begin{center}
\plotone{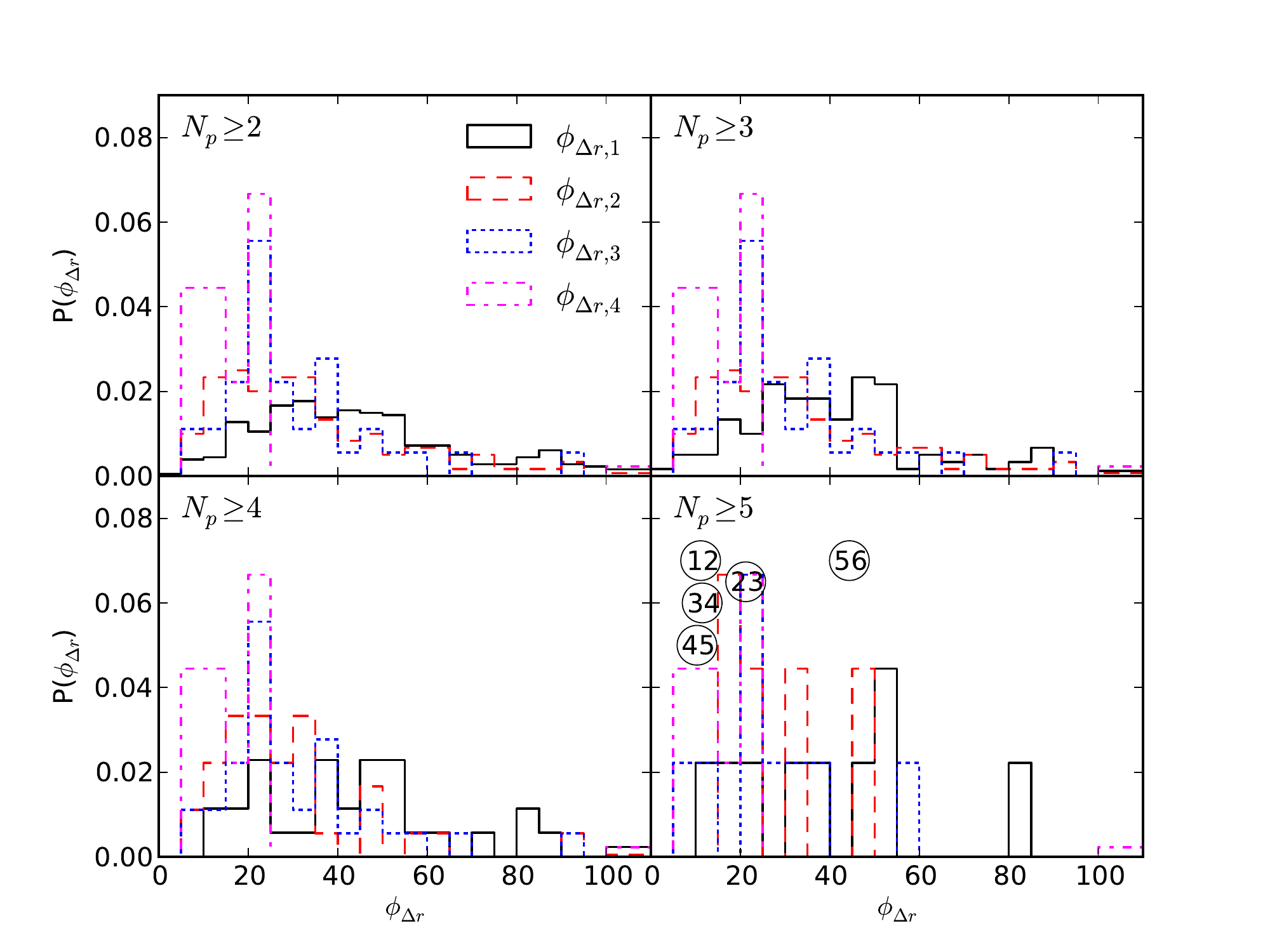}
\caption[Planetary spacings]{
Probability distribution of $\phideltar$ for KPC systems with
$N_p\geq$ 2 (top-left), 3 (top-right), 4 (bottom-left), and 5
(bottom-right). The $\phideltar$ values for the 6-planet system are
shown with inscribed circles in the bottom-right panel. 
Solid (black), dashed (red), dotted (blue), and dash-dot (magenta)
lines show $\phideltarone$ (separation between planets 1--2), to
$\phideltarfour$ (separation between planets 4--5), respectively.
Planets are indexed with increasing distance from the star.  
}
\label{fig:spacing}
\end{center}
\end{figure*}

Finally we consider orbital spacings between adjacent planets via
$\phideltari\equiv\deltari/R_{H,i}$, where $\deltar_i=r_{i+1}-r_{i}$
and $R_{H,i}$ is the Hill radius of the inner planet of the pair.  The
distributions of the large KPC sample are shown in
Figure~\ref{fig:spacing}, with a broad distribution peaking at
$\phideltar\sim 20$--50. The values for the TTV systems are listed in
Table~1 and are similar, with $\phideltar\gtrsim10$ and 4 of the 12
values clustered at $\phideltar\simeq14$. Thus $\phideltar$ is
typically at least several times greater than our fiducial value of
$\phi_H=3$ for gap opening, consistent with our theoretical
expectations that the spacing is determined not via dynamical
stability considerations (through $\phi_H R_H$) but via retreat of the
dead zone and associated location of the pressure maximum. However, it
is also possible that these spacings may be influenced by migration.

If $\phideltar$ is set by dead zone retreat one may expect greater
relative change immediately after formation of the first planet, since
this is the first gap-opening episode in the disk. Comparing
$\phideltar$ distributions in systems with $N_p\geq3$, $4$, $5$
planets (minimal detection bias is expected for interior planet
locations), indeed $\phideltarone$ tends to be larger than
$\phideltartwo$ and $\phideltarthree$. For $N_p\geq3$-sample, the KS
test gives $9\times10^{-5}$ probability that
($\phideltarone,\phideltartwo$) are drawn from the same
distribution. Equivalent probabilities for $N_p\geq4$-sample for
($\phideltarone,\phideltartwo$), ($\phideltarone,\phideltarthree$),
($\phideltartwo,\phideltarthree$) are $2\times10^{-4}$,
$5\times10^{-4}$, 0.8, respectively.

\section{Discussion and Summary}\label{S:summary}

We have presented a simple theoretical model of ``inside-out" planet
formation: pebbles form and drift to the inner disk; they accumulate
and dominate in a ring at the pressure maximum associated with the
inner dead zone boundary; a $\sim1\:M_\Earth$ planet forms, perhaps
initiated by gravitational instability of the ring;
inward Type I migration may bring the planet inside the MRI-active
region, allowing a new ring and planet to form at the dead zone
boundary; under certain conditions a planet may form that is massive
enough
to isolate itself from the disk by opening a deep gap; more typically,
and if Type I migration is inefficient, gap-opening would require the
planet to accrete additional mass (pebbles and/or gas); a variety of
mean planetary densities can arise, depending on the relative
importance of residual gas accretion; gap opening allows greater X-ray
penetration and the dead zone retreats; for a dead zone boundary set
by thermal ionization, a simple gradual reduction in accretion rate
would also lead to dead zone retreat; planet formation proceeds
sequentially, one at a time, from a series of retreating pebble rings,
as long as the supply of pebbles is maintained from the outer disk.

The {\it Kepler} STIPs planetary masses and relative orbital spacings
are consistent with expectations from this simple theoretical model,
for typical disk accretion rates $\sim10^{-9}\:\smyr$.  The
observed $\Mpl-r$ relationship agrees with the theoretical expectation,
although more data are needed to improve this test. Observed
dispersion of this relation within individual systems may indicate
accretion rate variability by factors of several during planet
formation.

Investigation of this model can be improved in several ways, including
(1) a more accurate calculation of disk structure that allows for
realistic opacity variations and heating from the central star; (2) an
estimate of the dead zone inner boundary involving an explicit
calculation of the ionization fraction; (3) a dynamical model that
tracks pebble formation, growth, and radial drift \citep[potentially subject
to secular instabilities leading to planetesimal formation,][]{2000Icar..148..537G} 
to form a pebble ring; (4) a dynamical model for planet
formation from such a pebble ring, which dominates the local mass
surface density of the disk, and especially its propensity to form a
single massive planet \citep[cf.][who investigated formation of clumps
  of solids from more gas-rich initial conditions mediated by
  hydrodynamic streaming instabilities and
  vortices]{2007Natur.448.1022J,2009ApJ...704L..75J,2010ApJ...722L.220B,2010ApJ...722.1437B};
(5) numerical investigation of migration, subsequent accretion and
gap-opening to the isolation mass
\citep[e.g.,][]{2013ApJ...768..143Z}, and resulting dead zone retreat.
Improved observational tests require better measurements of planetary
masses and densities. The dispersion in orbital inclination angles may
provide additional constraints that can help distinguish inside-out
planet formation from other formation models, such as formation from
an inner enriched disk or outer-disk formation followed by
long-distance migration.

The model of inside-out planet formation requires a sufficiently
high rate of supply of pebbles to the inner disk. The observed
diversity of planetary system architectures, from STIPs to that of our
own Solar System, may result from variations in both the efficiency
with which pebbles form and, once formed, their
ability to drift radially inwards through the disk without
interruption. Formation rate of pebbles is potentially related to the temperature
structure and the prevalence of icy dust grain mantles in the
bulk of the disk during the late stages of star formation. 
Future studies of the processes that lead to variation in inner disk
pebble supply rate are also needed.

\begin{deluxetable*}{ccccccccc}
\tabletypesize{\footnotesize}
\tablecolumns{8}
\tablewidth{0pt}
\tablecaption{KPC systems with direct mass measurements. }
\tablehead{
	  \colhead{Planet\tablenotemark{a}} &
           \colhead{$R_{\rm{pl}}$} &
           \colhead{$M_{\rm{pl}}$} &
           \colhead{$\rho_{\rm{pl}}$} &
           \colhead{$r$} & 
           \colhead{$\phideltar$\tablenotemark{b}} &
           \colhead{$k_M$\tablenotemark{c}} &
           \colhead{$k_M$\tablenotemark{d}} \\
           \colhead{Name} &
           \colhead{($R_\Earth$)} &
           \colhead{($M_\Earth$)} &
           \colhead{($\gccc$)} &
           \colhead{(AU)} & 
           \colhead{} &
           \colhead{Adjacent Pairs} &
           \colhead{System} \\
}
\startdata
Kepler-9b & $9.22\pm0.8$ & $80\pm4$ & $0.524 \pm 0.132$ & $0.140\pm0.001$ & $14\pm1$ & -0.8 & -0.8 \\
Kepler-9c & $9.01\pm0.7$ & $54\pm4$ & $0.383 \pm 0.098$ & $0.225\pm0.001$ & - & - & - \\
\hline
Kepler-11b & $1.8\pm0.02$ & $1.9^{+1.4}_{-1.0}$ & $1.77^{+1.29}_{-0.94}$ & $0.091\pm0.001$ & $14\pm5$ & 2.6 & 0.5 \\
Kepler-11c & $2.87^{+0.01}_{-0.02}$ & $2.9^{+2.9}_{-1.6}$ & $0.68^{+0.68}_{-0.36}$ & $0.107\pm0.001$ & $31\pm10$ & 2.5 & - \\
Kepler-11d & $3.11\pm0.02$ & $7.3^{+0.8}_{-1.5}$ & $1.33^{+0.15}_{-0.28}$ & $0.155\pm0.001$ & $13\pm2$ & 0.4 & - \\
Kepler-11e & $4.18\pm0.02$ & $8.0^{+1.5}_{-2.1}$ & $0.60^{+0.12}_{-0.16}$ & $0.195^{+0.002}_{-0.001}$ & $14\pm2$ & -5.6 & - \\
Kepler-11f & $2.48^{+0.02}_{-0.03}$ & $2.0^{+0.8}_{-0.9}$ & $0.73^{+0.30}_{-0.34}$ & $0.250\pm0.002$ & - & - & - \\
\hline
Kepler-18b & $2.0\pm0.1$ & $6.9\pm3.4$ & $4.9\pm2.4$ & $0.0447\pm0.0006$ & $35\pm9$ & 1.8 & 0.9 \\
Kepler-18c & $5.49\pm0.26$ & $17.3\pm1.9$ & $0.59\pm0.07$ & $0.0752\pm0.0011$ & $21\pm3$ & -0.1 & - \\
Kepler-18d & $6.98\pm0.33$ & $16.4\pm1.4$ & $0.27\pm0.03$ & $0.1172\pm0.0017$ & - & - & - \\
\hline
Kepler-20b & $1.91^{+0.12}_{-0.21}$ & $8.7^{+2.1}_{-2.2}$ & $6.5^{+2.0}_{-2.7}$ & $0.04537^{+0.00054}_{-0.00060}$ & $49\pm7$ & 0.8 & 0.4 \\
Kepler-20c & $3.07^{+0.20}_{-0.31}$ & $16.1^{+3.3}_{-3.7}$ & $2.91^{+0.85}_{-1.08}$ & $0.0930\pm0.0011$ & $104\pm12$ & - & - \\
Kepler-20d & $2.75^{+0.17}_{-0.30}$ & $<20$ & $<4.07$ & $0.3453^{+0.0041}_{-0.0046}$ & - & - & - \\
\hline
Kepler-36b & $1.486\pm0.035$ & $4.45^{+0.33}_{-0.27}$ & $7.46^{+0.74}_{-0.59}$ & $0.1153\pm0.0015$ & $7\pm2$ & 5.6 & 5.6 \\
Kepler-36c & $3.679\pm0.054$ & $8.08^{+0.60}_{-0.46}$ & $0.89^{+0.07}_{-0.05}$ & $0.1283\pm0.0016$ & - & - & - \\
\hline
Kepler-68b & $2.31^{+0.06}_{-0.09}$ & $8.3^{+2.2}_{-2.4}$ & $3.32^{+0.86}_{-0.98}$ & $0.06170\pm0.00056$ & $23\pm4$ & -1.4 & -0.6 \\
Kepler-68c & $0.953^{+0.037}_{-0.042}$ & $4.8^{+2.5}_{-3.6}$ & $28^{+13}_{-23}$ & $0.09059\pm0.00082$ & $878\pm228$ & -0.6 & - \\
Kepler-68d & - & $0.947\pm0.035$\tablenotemark{e} & & $1.4\pm0.03$ & - & - & - \\
\enddata
\tablenotetext{a}{Data for Kepler-9,11,18,20,36,68 from \citet{2010Sci...330...51H,2013arXiv1303.0227L,2011ApJS..197....7C,2012ApJ...749...15G,2012Sci...337..556C,2013ApJ...766...40G}, respectively. }
\tablenotetext{b}{$\phideltar=(r_{i+1}-r_i)/R_{H,i}$.}
\tablenotetext{c}{$M_{\rm{pl}} \propto r^{k_M}$ fitted for adjacent pairs. }
\tablenotetext{d}{$M_{\rm{pl}} \propto r^{k_M}$ fitted for whole system.}
\tablenotetext{e}{Radial velocity measurement of $M_{\rm{pl}}\sin i$. } 
\label{tab:list}
\end{deluxetable*}

\acknowledgments We thank Aaron Boley, Eric Ford, Brad Hansen, Anders
Johansen, Greg Laughlin, Subu Mohanty, Ralph Pudritz, Andrew Youdin,
Yichen Zhang, Jeremy Goodman, Zhaohuan Zhu, Roman Rafikov, Scott
Tremaine, and Leonardo Testi for helpful discussions. SC acknowledges
NASA grants NNX08AR04G, NNX12AF73G and the UF Theory Postdoctoral
Fellowship.  JCT acknowledges NASA grants ATP09-0094, ADAP10-0110.


\end{document}